\documentstyle[twoside,fleqn,espcrc2,epsf,axodraw]{article}
\begin{document}
\title{Heavy Quarkonium Physics beyond the Next-to-Next-to-Leading
Order of NRQCD
\thanks{Talk presented at the High Energy
Physics International Conference on Quantum Chromodynamics
(QCD-00), Montpellier, France, 6-12 July 2000.}
}

\author{Alexander A. Penin
\address{II.\ Institut f\"ur Theoretische Physik, Universit\"at Hamburg, 
Luruper Chaussee 149, D-22761 Hamburg, Germany}\thanks{Permanent address:
Institute for Nuclear Research,  60th October Anniversary Pr., 
7a, Moscow 117312, Russia}}

\begin{abstract}
In this talk I briefly review the recent progress in  
calculation of the high 
order corrections to the parameters of the  nonrelativistic
heavy quark-antiquark system in the effective theory 
approach.  
\end{abstract}

\maketitle

\section{INTRODUCTION}

The theoretical study of the nonrelativistic  
heavy quark-antiquark system \cite{AppPol}  and its application
to bottomonium \cite{NSVZ} and toponium   \cite{FadKho}
physics is of special interest because it relies 
entirely on the  first principles of QCD. 
The system in principle allows for a perturbative treatment
with the nonperturbative effects  being well under control
and with no crucial model dependence. This  makes the heavy
quarkonium to be  an ideal
place to determine the fundamental QCD parameters
such as the heavy quark mass $m_q$ and the strong coupling
constant $\alpha_s$.  Recently an essential progress 
has been made in the theoretical 
investigation of the nonrelativistic heavy quark dynamics
based on  the effective theory approach  \cite{CasLep}.
The analytical results for the main parameters of
the nonrelativistic heavy quark-antiquark system
are now available up to next-to-next-to-leading order (NNLO) 
in the strong coupling  constant and the heavy quark velocity $\beta$
\cite{Pet,CzaMel,PinYnd,KPP,HoaTeu1,MelYel1,PenPiv1,Hoa1,MelYel2,PenPiv2,PenPiv3,BSS,NOS,HoaTeu2,BenSin,Hoa2,gang}.
The NNLO corrections have turned out to be so sizeable that it appears to be
indispensable also to gain control over the next-to-next-to-next-to-leading
order (N$^3$LO) both in regard of phenomenological applications and in order 
to understand the structure and the peculiarities of the nonrelativistic
expansion. In this talk  we review 
the first steps in this direction and consider two 
particular classes of N$^3$LO corrections: (i) 
the retardation effects arising from the 
emission and absorption of dynamical ultrasoft gluons by the heavy 
quarks and (ii) the  corrections enhanced by a  power of 
$\ln(1/\alpha_s)$ which are not generated by the renormalization group
(RG) running of $\alpha_s$.

\section{EFFECTIVE THEORY OF NON- RELATIVISTIC HEAVY QUARKS}
In this section we briefly outline the 
effective theory approach to the nonrelativistic
heavy quark dynamics.
Let us consider the near threshold behavior of the heavy quark
vacuum polarization function  
\[
\left(q_\mu q_\nu-g_{\mu \nu}q^2\right)\Pi(q^2)
\]
\begin{equation}
=i\int d^4xe^{iq\cdot x}\langle 0|Tj_{\mu}(x)j_{\nu}(0)|0\rangle\,,
\end{equation}
where
$j_\mu=\bar q\gamma_\mu q$ is the heavy quark vector current.
Its imaginary part is related to the normalized cross section of $q\bar q$
production in the photon mediated
$e^+e^-$ annihilation at energy $s=q^2$ 
\begin{equation}
R(s)={\sigma(e^+e^-\rightarrow q\bar q)\over
\sigma(e^+e^-\rightarrow\mu^+\mu^-)}\,,
\end{equation}
by
\begin{equation}
R(s)={12\pi Q_q^2}{\rm Im}\Pi(s+i\epsilon)\, ,
\end{equation}
where $Q_q$ is the fractional charge of quark $q$.
The behavior of the vacuum polarization function near-below  
the threshold  $s= 4m^2_q$ determines also the masses and leptonic 
widths  of the perturbative heavy quarkonium bound states.
 
Near the  threshold the heavy quarks are nonrelativistic
so that one may consider
the quark velocity $\beta=\sqrt{1-4m_q^2/s}$  as a small parameter.
An expansion in $\beta$ may be performed directly in the Lagrangian of QCD by
using the framework of effective field theory.
In the nonrelativistic problem there are four different scales \cite{BenSmi}:
(i) the hard scale (energy and momentum scale like $m_q$);
(ii) the soft scale (energy and momentum scale like $\beta m_q$);
(iii) the potential scale (energy scales like $\beta^2m_q$ while 
momentum scales like $\beta m_q$); and
(iv) the ultrasoft scale (energy and momentum scale like $\beta^2m_q$).
The ultrasoft scale is only relevant for gluons.  
By integrating out the hard scale of QCD one arrives at the effective theory
of nonrelativistic QCD  (NRQCD) \cite{CasLep}. 
If one also integrates out the soft scale and the potential gluons one 
obtains the effective theory of potential NRQCD (pNRQCD) 
 which contains potential quarks and
ultrasoft gluons as active particles \cite{PinSot1}.
The dynamics of the quarks is governed by the effective nonrelativistic 
Schr\"odinger equation and by their interaction with the ultrasoft gluons.
To get a regular perturbative expansion within pNRQCD  this interaction should
be expanded in multipoles.
The corrections from harder scales are contained in the Wilson coefficients
leading to an expansion in $\alpha_s$ as well as in the higher-dimensional
operators of the nonrelativistic Hamiltonian corresponding to an expansion in
$\beta$.
The nonrelativistic expansion in $\alpha_s$ and $\beta$ provides us with the
following representation of the heavy quark vacuum polarization function near
threshold
\begin{equation}
\Pi(E)={N_c\over2m_q^2}C(\alpha_s)G(0,0,E)+\ldots\,,
\end{equation}
where $E=\sqrt{s}-2m_q$ is the $q\bar q$ energy counted from the threshold,
$C(\alpha_s)$ is the square of the hard renormalization 
coefficient of the  nonrelativistic vector current 
and the ellipsis stands for the higher-order
terms in $\beta$.
$G({\bf x},{\bf y},E)$ is the nonrelativistic Green function which sums up
the $(\alpha_s/\beta)^n$ terms singular near the threshold.
It is determined by the Schr\"odinger equation which describes the the
propagation of the nonrelativistic quark-antiquark pair in pNRQCD
\begin{equation}
\left({\cal H}-E\right)G({\bf x},{\bf y},E)
=\delta^{(3)}({\bf x}-{\bf y}),
\end{equation}
where ${\cal H}$ is the nonrelativistic Hamiltonian defined by
\begin{equation}
{\cal H}=-{{\bf \partial}_{\bf x}^2\over m_q}+V(x)+\ldots,\quad
V(x)=V_C(x)+\ldots\,.
\label{Schr}
\end{equation}
Here $V_C(x)=-C_F\alpha_s /x$ is the Coulomb potential, $C_F=4/3$ is the
eigenvalue of the quadratic Casimir operator of the fundamental representation
of the color group, $x=|{\bf x}|$ and the ellipses stands for the
higher-order terms in $\alpha_s$ and  $\beta$.
The Green function has the spectral representation
\[
G({\bf x},{\bf y},E)
=\sum_{n=1}^\infty{\psi^*_n({\bf x})\psi_n({\bf y})\over E_n-E}
\]
\begin{equation}
+\int_0^\infty{d^3k\over(2\pi)^3}\,
{\psi^{*}_{\bf k}({\bf x})\psi_{\bf k}({\bf y})\over k^2/m_q-E}\,,
\label{spectr}
\end{equation}
where $\psi_m$ and $\psi_{\bf k}$ are the wave functions of the $q\bar q$
bound and continuum states respectively.

Below the threshold the  vacuum polarization function of a
stable heavy quark is  determined by the bound state parameters. 
For the leading order Coulomb (C) Green function the energy levels and wave
functions at the origin read
\begin{equation}
E^C_n=-{\lambda_s^2\over m_qn^2},\qquad
\left|\psi_n^C(0)\right|^2={\lambda_s^3\over \pi n^3}\, ,
\end{equation}
where $\lambda_s=\alpha_sC_Fm_q/2$.
Note that for the study of the bound-state parameters we have
$\beta\approx\alpha_s$ so that we are only dealing with one expansion
parameter.

The current status of the theoretical research 
can be summarized  as follows.
 The effective nonrelativistic Hamiltonian is known 
in the NNLO approximation including the two-loop corrections 
to the static potential \cite{Pet} and the corresponding
corrections to the Green function have been obtained
in analytical form in \cite{PinYnd,KPP,HoaTeu1,MelYel1,PenPiv1,Hoa1,MelYel2}.  
The ${\cal O}(\alpha_s^2)$ contribution
to the hard renormalization coefficient has been computed
in  \cite{CzaMel}.
Thus the complete NNLO analytical expression for the
nonrelativistic heavy quark
vacuum polarization function  
is now available.  
In the next section we present the recent results of the calculation of
some N$^3$LO  order corrections to the  heavy quarkonium parameters.

\section{HEAVY QUARKONIUM PARAMETERS BEYOND THE NNLO
OF  NRQCD}
\subsection{Retardation effects}
Let us start with a few general remarks concerning the structure of
higher-order corrections in pNRQCD.
At next-to-leading order (NLO) 
the only source of corrections is the perturbative
corrections to the hard renormalization coefficient and 
the static potential.
In NNLO the higher-dimensional operators start to contribute.
In N$^3$LO the retardation effects which cannot be described by 
operators of instantaneous interaction enter the game.
The  retardation effects  are induced by 
the emission and absorption of virtual ultrasoft 
gluons which ``feel''  the binding effects
of quark-antiquark interaction  \cite{KniPen1,BPSV1}. 
This constitutes a genuinely new feature which is absent in
NLO and NNLO and is supposed to be the last qualitatively 
new source of the corrections. 

The leading retardation effects which are under consideration here arise
from the chromoelectric dipole interaction 
$g_s({\bf r_q}-{\bf r_{\bar q}})\cdot{\bf E}$
of heavy quarkonium with virtual
ultrasoft gluons.
The corresponding diagram is shown in Fig.~1a.
\begin{figure}[t]
   \vspace{-3.1cm}
   \epsfysize=19.9cm
   \epsfxsize=13.1cm
   \centerline{\epsffile{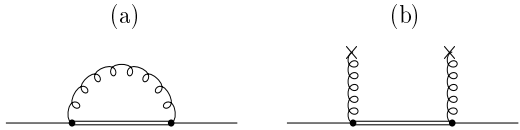}}
   \vspace{-15.3cm}
\caption[dummy]{\label{fig1}\small 
(a) Feynman diagram giving rise to the ultrasoft contribution at
N$^3$LO.
The single and double lines stand for the singlet and octet Green functions,
respectively, the wavy line represents the ultrasoft-gluon propagator in the 
Coulomb gauge and the vertices correspond to the chromoelectric dipole
interaction.
(b)  Feynman diagram giving rise to the leading nonperturbative
corrections. The crossed wavy lines represent the 
vacuum fluctuations of the gluonic field.} 
\vspace{-0.6cm}
\end{figure}
An interesting distinction from the similar QED
process is that after emitting
the ultrasoft gluon the quark-antiquark
pair converts into the color octet sate 
with repulsive Coulomb potential  $V^o_C(x)=(C_A/2-C_F)\alpha_s /x$ 
where $C_A=3$ is the eigenvalue of the quadratic Casimir operator of the
adjoint representation of the color group.
As a consequence
the intermediate color octet states  belong
to the continuum part of the spectrum only.

The result for this ultrasoft 
contribution diverges in the ultraviolet (UV)
region. This divergence is spurious.
It arises in the process of scale separation due to the use of pNRQCD
perturbation theory at short distances where it is inapplicable. 
We use the dimensional regularization (DR) 
with $d=4-2\epsilon$ space-time dimensions
to handle the UV and infrared (IR) divergences of the effective 
theory which are of the form $1/\epsilon^n$
($n=1,2,\ldots$) as $\epsilon\to0$ \cite{BenSmi,PinSot2,CMY}.
Compared to the ``traditional'' NRQCD approach endowed with an explicit
momentum cutoff and a fictitious photon mass to regulate the ultraviolet and
infrared behavior \cite{CasLep} this scheme has the advantage
that the contributions from the different scales are matched automatically. 
In the total N$^3$LO result the poles in $1/\epsilon$ in the
ultrasoft contribution are canceled by the 
infrared poles coming from the hard and soft scale corrections. 
However since these corrections are still unknown we subtract the divergent
part according to the $\overline{\rm MS}$ scheme.
This means that the same scheme must be used for the calculation of the hard
and soft scale corrections.
As a consequence the partial result for the ultrasoft contribution depends on
the auxiliary ultrasoft scale $\mu_{us}$ which drops out in the total
result.

The corresponding corrections to the Coulomb 
energy levels and the wave functions at the origin 
\[
E_n=E^C_n+\Delta E_n\,,
\]
\begin{equation}
|\psi_n(0)|^2=\left|\psi^C_n(0)\right|^2\left(1+\Delta\psi_n^2\right)
\end{equation}
read \cite{KniPen1}
\[
\Delta E_n=-{2\alpha_s^3\over 3\pi}E^C_n\left\{
\left[{1\over 4}C_A^3+{2\over n}C_A^2C_F\right.\right.
\]
\begin{equation}
\left.
+\left({6\over n}-{1\over n^2}\right)C_AC_F^2+
{4\over n}C_F^3\right]
\label{enret}
\end{equation}
\[
\times\left.
\left(\ln{\mu_{us}\over E^C_1}+{5\over6}-\ln{2}\right)+C_F^3L^E_n\right\}\,,
\]
\[
\Delta\psi^2_n=-{2\alpha_s^3\over\pi}\left[
\left({1\over4}C_A^2C_F+{1\over n^2}C_AC_F^2+
{1\over n^2}C_F^3\right)\right.
\]
\begin{equation}
\left.\times\left(\ln{\mu_{us}\over E^C_1}+{1\over 2}-\ln{2}\right)
+C_F^3L^\psi_n\right]\,.
\label{wfret}
\end{equation}
Here $L^E_n$ and $L^\psi_n$ are the QCD analogs of the 
famous QED ``Bethe logarithms''.
They  represent a pure
retardation effect which cannot be reduced to the instantaneous
interaction contribution.
For $n=1,~2,~3$  their numerical  values are \cite{KniPen1}
\begin{equation}
\begin{array}{lll}
L^E_1=-81.5379\,,\qquad&L^\psi_1=-5.7675\,,\\
L^E_2=-37.6710\,,\qquad&L^\psi_2=\hspace{2.8mm}0.7340\,,\\
L^E_3=-22.4818\,,\qquad&L^\psi_3=\hspace{2.8mm}2.2326\,.\\
\end{array}
\end{equation}

\subsection{Non-RG  logarithmic corrections}
The origin of the logarithmic corrections 
which  are not generated by the  
RG running of $\alpha_s$
is the presence of several scales
in the nonrelativistic dynamics.
The logarithmic integration over a loop momentum between different scales 
yields a power of $\ln(1/\beta)$
and  in the  approximately  Coulomb system we have 
$\beta\propto\alpha_s$. 
These corrections can always be associated with
the divergences of the effective theory 
\cite{BPSV2,KniPen2,KniPen3}.
By contrast the RG logarithms are well known and may be resummed by an
appropriate scale choice. 

In the heavy quarkonium 
the non-RG  logarithms first arise in NNLO
corrections to the wave functions at the
origin. They  are generated by the following ${\cal O}(\beta^2)$
operators in the effective
nonrelativistic Hamiltonian:
\[
-{C_FC_A\alpha_s^2\over2m_qx^2}
+{C_F\alpha_s\over2m_q^2}\left\{{\bf\partial}_{\bf x}^2,{1\over x}\right\}
\]
\begin{equation}
+\left(1+{4\over3}{\bf S}^2\right){\pi C_F\alpha_s\over m_q^2}\delta({\bf x})
-{{\bf\partial}_{\bf x}^4\over4m_q^3}\, ,
\label{delHam1}
\end{equation}
where $\bf S$ represents the spin of the
quark-antiquark system and $\{.,.\}$ denotes the anticommutator.
The corresponding corrections to the wave functions
are proportional to the  Coulomb Green function at the origin
which is UV divergent and in DR is of the following form
\begin{equation}
G_C(0,0,E)
={m_q\lambda_s\over 4\pi}\left({1\over \epsilon}+
\ln{-\mu^2\over m_qE}
+\ldots\right)\, ,
\label{gf}
\end{equation}
where  the ellipsis stands for the nonlogarithmic contribution.
The pole term in Eq.~(\ref{gf}) is canceled by the 
${\cal O}(\alpha_s^2)$
IR pole of the hard matching coefficient $C(\alpha_s)$.
Thus the scale $\mu$ in the logarithm is to be identified with 
the hard scale $m_q$ 
and the sought corrections for the spin one states 
which are of the main interest
read
\cite{PenPiv1,MelYel2} 
\begin{equation}
\Delta\psi_n^2(0)={C_F\alpha_s^2}
\left({2\over 3}C_F+C_A\right)\ln{1\over\alpha_s}\, .
\label{wfll1}
\end{equation}
The residual operators in the NNLO effective nonrelativistic Hamiltonian which
are not contained in Eq.~(\ref{delHam1}) correspond to the purely perturbative 
corrections to the static Coulomb potential.
The corresponding corrections to the wave functions at the origin
\cite{PenPiv1,MelYel2,PenPiv2}  contain RG logarithms of the form
$\alpha_s^2\ln^m(\mu/\alpha_sm_q)$ ($m=1,2$) which vanish for
$\mu =\alpha_sm_q$.
This also holds in N$^3$LO for the RG logarithms of the form
$\alpha_s^3\ln^m(\mu/\alpha_sm_q)$ ($m=1,2,3$) because the ultrasoft effects
enter the stage only in N$^3$LO so that the corresponding running of the
strong coupling constant at the ultrasoft scale only becomes relevant in
N$^4$LO. An important point here is that starting from NNLO the hard matching
coefficient $C(\alpha_s)$ receives a nonvanishing anomalous dimension.
Therefore starting from N$^3$LO the running of $\alpha_s$ in $C(\alpha_s)$
should be taken into account to match the scale dependence
of the wave functions at $\mu=\alpha_sm_q$.

In N$^3$LO the non-RG leading logarithmic corrections 
are produced by the one-loop renormalization of the 
${\cal O}(\beta^2)$ operators. 
In dimensional regularization the pole part of the correction is
\[
{1\over 2\epsilon}\,{C_F\alpha_s\over\pi}
\left\{-{C_A^3\alpha_s^3\over 12x}-\left({4\over 3}C_F
+{2\over 3}C_A\right){C_A\alpha_s^2\over m_qx^2}\right.
\]
\[
+{2\over 3}\,{C_A\alpha_s\over m_q^2}
\left\{{\bf\partial}^2_{\bf x},{1\over x}\right\}
-\left({16\over3}C_F-{8\over3}C_A\right){\pi\alpha_s\over m_q^2}
\delta({\bf x})
\]
\begin{equation}
+\left.
\left[{2\over3}C_F+\left({17\over3}-{7\over3}{\bf S}^2\right)C_A\right]
{\pi\alpha_s\over m_q^2}\delta({\bf x})\right\}\, ,
\label{delHam2}
\end{equation}
where the first three terms contained within the parentheses represent the IR
divergence while the fourth one embodies the UV divergence of the potential.
The IR poles are canceled by the ultrasoft contribution with the
characteristic scale $\alpha_s^2m_q$ and may be read off from
\cite{KniPen1,BPSV2} while the UV poles are canceled by the IR poles of
the hard coefficients and may be extracted from \cite{Man,PinSot3}.  
Thus the divergences of the effective theory
endow the operators in the nonrelativistic
Hamiltonian with anomalous dimensions. This results 
in the logarithmic corrections to the spin one $l=0$  energy levels
\cite{BPSV2,KniPen2}
\[
\Delta E_n=-E_n{\alpha_s^3\over\pi}\left\{{3\over n}C_F^3
+\left[{9\over 2n}-{2\over3n^2}\right]C_F^2C_A\right.
\]
\begin{equation}
\left.
+{4\over3n}C_FC_A^2+{1\over6}C_A^3\right\}
\ln{1\over\alpha_s}\, .
\label{enll}
\end{equation}
Note that the IR poles of Eq.~(\ref{delHam2}) introduce a factor
$\ln(E_1^C/\lambda_s)\approx\ln\alpha_s$ while the UV poles contribute a
factor $\ln(m_q/\lambda_s)\approx\ln(1/\alpha_s)$. 

Some of the operators with singular coefficients in Eq.~(\ref{delHam2})
also result in the contributions to the wave functions 
which is proportional to
the singular Coulomb Green function at the origin.
The overlapping logarithmic divergences
lead to the double pole in  $1/\epsilon$ and therefore
to the double logarithmic contributions 
which for the spin one states read \cite{KniPen2}
\[
\Delta\psi_n^2(0)=-{C_F\alpha_s^3\over\pi}
\left[{3\over2}C_F^2+{9\over 4}C_FC_A\right.
\]
\begin{equation}
\left.
+{2\over 3}C_A^2\right]\ln^2{1\over\alpha_s}\, .
\label{wfll2}
\end{equation}

The calculation of the subleading single logarithmic contributions 
to the  wave functions is much more involved and the 
complete result for QCD bound states is not available.
However two important parts of these corrections are known.
The Abelian part of the correction to the
ground sate wave function can be read of the 
positronium lifetime analysis \cite{KniPen3}.
Not that one has to exclude the one photon annihilation
contribution from the positronium result because
this is absent in QCD due to the color conservation. 
The  trivial non-Abelian contribution 
originates from the interference
between the non-Abelian part of Eq.~(\ref{wfll1})
and the ${\cal O}(\alpha_s)$ term in the hard renormalization 
coefficient. The total known   single logarithmic contribution
to the spin one  ground state wave function reads
\[
\Delta\psi_n^2(0)={C^2_F\alpha_s^3\over\pi}
\left[\left({7\over 90} -8\ln{2}\right)C_F\right.
\]
\begin{equation}
-4C_A\bigg]\ln{1\over\alpha_s}\, .
\label{wfsl}
\end{equation}
Several attempts have been made to sum up
the high order RG and non-RG logarithmic corrections 
\cite{BSS,LMR}. 
The problem however is not completely solved yet.

\section{PHENOMENOLOGICAL APPLICATIONS}
\subsection{Top quark phenomenology}
The relatively large 
electroweak  top quark width $\Gamma_t$ 
and characteristic scale of the 
nonrelativistic Coulomb dynamics 
$\alpha_s^2m_t$ are considerably larger than $\Lambda_{\rm QCD}$ 
and serve  as an effective infrared cutoff for long distance 
nonperturbative strong interaction effects.
This makes perturbative 
QCD applicable for the theoretical description of the threshold
top quark production. At the same time  numerically
$\Gamma_t\sim \alpha_s^2m_t$ and 
the Coulomb effects are not completely 
dumped by the non-zero top quark width 
that should be properly taken into account.

\begin{figure}[t]
   \vspace{-2.95cm}
   \epsfysize=19cm
   \epsfxsize=13.4cm
   \centerline{\epsffile{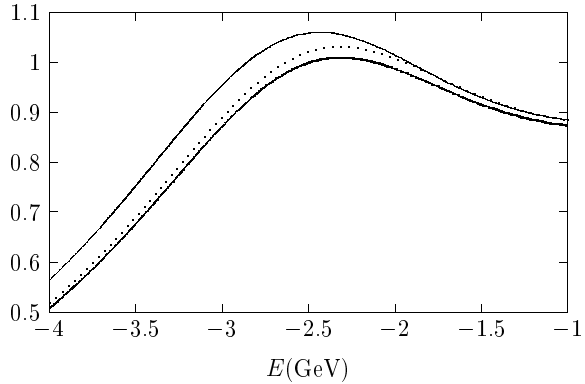}}
   \vspace{-11.5cm}
\caption[dummy]{\label{fig2}\small 
The normalized cross section $R(E)$ 
in NNLO (solid line)
and with the complete leading logarithmic 
and known part of the subleading logarithmic  
N$^3$LO corrections included (dotted line 
and bold solid line respectively). }
\vspace{-0.5cm}
\end{figure}

In order to analyze the significance of the N$^3$LO  logarithmic
corrections to the cross section $R(E)$ we start from the NNLO  
result of \cite{PenPiv3} and add the contributions from 
Eqs.~(\ref{enll},~\ref{wfll2}) and
Eq.~(\ref{wfsl}).
The results are plotted on Fig.~2. 
The input parameters are taken to be $\alpha_s(M_Z)=0.118$, $m_t=175$~GeV and
$\Gamma_t=1.43$~GeV.
The soft normalization scale of $\alpha_s$ 
in the nonrelativistic Coulomb problem is determined from the condition
$\mu_s=2\alpha_s(\mu_s)m_t$. The relatively large
soft normalization point is taken to improve
the convergence on the perturbative series for the
spectral density in NNLO.
The instability of the top quark is
implemented by the complex energy shift 
$E\to E+i\Gamma_t$ in Eq.~(\ref{Schr}).
This  accounts for the leading imaginary 
electroweak contribution to the  pNRQCD Hamiltonian.  
Numerically the  non-RG logarithmic corrections
are comparable to the contribution from the N$^3$LO RG logarithms 
which may be estimated from the renormalization scale 
dependence of the NNLO result.
The effect of the N$^3$LO logarithms is twofold.
The normalization of the cross section is reduced by about 10\% around the $1S$
peak  and the energy gap between the $1S$ peak and the nominal 
threshold is decreased by roughly $10\%$.
This  partially compensates the effect of huge NNLO corrections.

Since the ultrasoft corrections are scheme dependent
we do not include them into the numerical analysis.
However it is quite interesting to estimate their
numerical importance. 
If we take $\mu_{us}=\alpha_s^2m_t$ to cancel the
logarithms of $\alpha_s$ in Eqs.~(\ref{enret},~\ref{wfret}) the
ultrasoft correction to the $1S$  peak energy
is about $-200$~MeV and to the $1S$  peak normalization is 
only about $+2\%$ due to some cancellations.
Note that  one power of $\alpha_s$ in these equations
refers to the ultrasoft gluon interaction and should 
be evaluated at the ultrasoft scale
$\alpha_s^2m_t$
while the two residual powers of $\alpha_s$ originate from
the Coulomb Green function and should be evaluated at the soft scale
$\alpha_sm_t$. 

The stability of the perturbation theory 
for  the  $1S$ peak energy (but not for its normalization!) up to
NNLO  can be manifestly improved by using
an infrared safe mass parameter instead of
the pole mass in the analysis \cite{MelYel2,BSS,NOS,HoaTeu2,BenSin,Hoa2}.
We should note that the corrections under 
consideration are nor related to the RG running of
$\alpha_s$ and cannot be taken into account 
by the renormalon based mass redefinition.
On the other hand in some cases the behavior of the
perturbation perturbation theory can
be improved by using the direct relations  between
the  physical observables \cite{PenPiv1}.

\subsection{Bottom quark phenomenology}

In the case of bottom quark-antiquark production the nonperturbative effects
are much more significant and one is led to use the sum rule approach
\cite{NSVZ} to get them under control.
Specifically appealing quark-hadron duality one matches the theoretical 
results for the moments of the spectral density
\begin{equation}
{\cal M}_n=(4m_q^2)^n\int_0^\infty ds{R(s)\over s^{n+1}}\, .
\label{mom}
\end{equation}
For sufficiently large $n$ the moments are saturated by 
the near-threshold region. 
Then the main contribution to the experimental moments comes from the
$\Upsilon$ resonances which are measured with high precision.
On the other hand for $n$ of order 
${\cal O}\left(1/\alpha_s^2\right)$ the Coulomb
effects should be properly taken into account on the theoretical side. 
In order to analyze the N$^3$LO  logarithmic corrections to the
$\Upsilon$ sum rules we upgrade the NNLO result of \cite{PenPiv1} by
including Eqs.~(\ref{enll},~\ref{wfll2}) and
Eq.~(\ref{wfsl}).
We fix the strong coupling constant  and focus on the
determination of the bottom quark mass.
At present this seems to be the most interesting application of
the sum rules.
In the bottom-quark case $\alpha_s$ at the nominal
ultrasoft scale  seems to be too large 
for a reliable perturbative calculation.
Thus to be on the safe side we redefine  
the ultrasoft scale to be 
$3\alpha_s^2m_b$ so that in both cases 
$\alpha_s=0.34$. 
We find that the inclusion of the N$^3$LO  logarithms in the sum rules
leads to a reduction of the extracted $\overline{\rm MS}$ 
mass value by approximately 150~MeV for
moderate values of $n$, $5<n<15$.  
The result essentially depends on $\mu_s$ and  $\mu_{us}$
because the
scale dependence of the correction is compensated only by higher-order terms.
For example, for $\mu_s\sim\mu_{us}\sim m_b$ the perturbative
series for the moments converges much better
and the correction to the 
$\overline{\rm MS}$ mass
is only about $-30$~MeV. This result  implies that the unsertainty
of the  value $m_{\overline{\rm MS}}(m_{\overline{\rm MS}})\approx4.2$~GeV
obtained from NNLO analysis of $\Upsilon$
sum rules \cite{PenPiv1,MelYel2,BenSin,Hoa2} is  at least about $100$ MeV. 

In the case of the  $\Upsilon(1S)$ meson the local duality is expected to
work though the nonperturbative
effects are much more important
than in the sum rules approach. 
If we assume the  $\Upsilon(1S)$ meson to be a perturbative system
the results of Sect.~3 can be applied to compute the corrections
to its mass which is determined by the binding energy
and leptonic width which is determined by the wave function at the origin.
In this way we find that the N$^3$LO  logarithmic correction 
to the  $\Upsilon(1S)$ mass
is about $-170$~MeV and to its leptonic width  is 
about $-70\%$ (the leading and subleading logarithms are 
approximately equal).
The ultrasoft correction to the  $\Upsilon(1S)$ mass
is about $-110$~MeV and to its leptonic width  is 
about $-10\%$.

It is interesting to compare the perturbative ultrasoft contribution
to the leading nonperturbative contribution of the gluonic condensate  due to 
vacuum fluctuations of the gluonic field
at the scale $\Lambda_{\rm QCD}$ 
which is generated by the  similar diagram  with the broken gluon propagator 
shown in Fig.~1b  in order to conclude how 
``perturbative'' the heavy quarkonium is.
In the case of the  energy level the leading nonperturbative
contribution is given by \cite{VolLeu}
\begin{equation}
\Delta E_1=\frac{117m_q}{1275\lambda_s^4}
\left\langle\alpha_s G^a_{\mu\nu} G^{a\mu\nu}\right\rangle \, .
\end{equation}
Using the standard literature value
$\langle\alpha_sG^2\rangle\approx0.06$~GeV$^4$ we have
$\Delta E_1\approx 60$~MeV which is of the same scale as  the 
ultrasoft contribution.

\section{CONCLUSION}
In this talk we reviewed the  first steps towards the 
N$^3$LO analysis of the nonrelativistic 
heavy quark-antiquark system.
Two special classes of N$^3$LO contributions 
to the key
parameters of heavy quark-antiquark bound states
were considered, namely the non-RG corrections
enhanced by a  power of $\ln(1/\alpha_s)$ and 
the retardation effects. 
They are not related
to the RG
running of $\alpha_s$    and 
can be considered as typical 
representatives of the N$^3$LO corrections.

We  gave
the numerical estimates of the above corrections
for the  top and bottom quark  systems.
The corrections  
turn out to be  comparable to the NNLO ones and reach
10\% in magnitude even in the case of top 
where $\alpha_s\approx1/10$. For $\Upsilon(1S)$ meson
they are out of control. 
This tells us that the NRQCD  expansion is not a fast convergent
series for the physical value of the strong coupling constant.
It is highly desirable to complete the calculation of N$^3$LO
corrections to get more insight 
of the nonrelativistic heavy quark dynamics and  
the structure of the nonrelativistic perturbation theory.
Although there is no conceptual problem on the theoretical side
this analysis is extremely difficult
from the technical point of view and includes, for example, 
the calculation 
of the three-loop hard matching coefficient and the 
three-loop static potential.

Finally we would like to mention 
that the subleading  nonperturbative contributions \cite{Pin}
and charm quark mass effects \cite{EirSot}
which are relevant for the bottom quark physics  
but  numerically 
are not so important as the perturbative corrections.

\vspace{4mm}
\noindent
{\bf Acknowledgments}\\[1mm]
The author thanks the organizers of QCD-00 conference
for making this fruitful meeting.
This work is supported in part by 
the Bundesministerium f\"ur Bildung und Forschung
under Contract No.\ 05~HT9GUA~3
and  by the Volkswagen Foundation under
Contract No.\ I/73611.

\end{document}